\documentclass[12pt]{article}
\usepackage{epsfig}

\begin{document}

\title{\textbf{Innovation Success and Structural Change: An Abstract Agent Based
Study}\thanks{%
Supported by FCT Portugal project PDCT/EGE/60193/2004}}
\author{\textbf{Tanya Ara\'{u}jo}\thanks{
tanya@iseg.utl.pt, http:www.iseg.utl.pt/tanya} \\
ISEG, Universidade T\'{e}cnica de Lisboa (TULisbon) and\\
Research Unit onComplexity in Economics (UECE)\\
\textbf{R. Vilela Mendes}\thanks{%
vilela@cii.fc.ul.pt, http:label2.ist.utl.pt/vilela}\\
Centro de Matem\'{a}tica e Aplica\c{c}\~{o}es Fundamentais and\\
Universidade T\'{e}cnica de Lisboa}
\maketitle

\begin{abstract}
A model is developed to study the effectiveness of innovation and its impact
on structure creation and structure change on agent-based societies. The
abstract model that is developed is easily adapted to any particular field.
In any interacting environment, the agents receive something from the
environment (the other agents) in exchange for their effort and pay the
environment a certain amount of value for the fulfilling of their needs or
for the very price of existence in that environment. This is coded by two
bit strings and the dynamics of the exchange is based on the matching of
these strings to those of the other agents. Innovation is related to the
adaptation by the agents of their bit strings to improve some utility
function.
\end{abstract}


\section{Introduction}

Agent-based models are increasingly being used to model artificial
societies. Some of these models fall in the field of biological sciences and
a very important part of them deal with economical problems (\cite{Kirman} 
\cite{LeBaron} \cite{Grossman} \cite{Nelson} \cite{Blume} \cite{Danilov} and 
\cite{Dawid}). Economical, ecological and social environments share as a
common feature the fact that the agents operating in these environments
spend a large amount of their time trying to maximize some kind of actual or
perceived utility function, related either to profit, to food, to
reproduction or to comfort and power. It so happens that many times the
improvement of one agent's utility is made at the expense (or causes) the
decrease of the other agents utilities. A general concept that is attached
to this improvement struggle is the idea of \textit{innovation}.

In the economy, innovation may be concerned with the identification of new
markets \cite{Gima} \cite{Johne}, with the development of new products \cite
{Souder} \cite{Arthur} to capture a higher market share or with the
improvement of the production processes to increase profits. In ecology,
innovation concerns better ways to achieve security or food intake or
reproduction chance and, in the social realm, all of the above economical
and biological drives plus a few other less survival-oriented needs. In all
cases, innovation aims at finding strategies to better deal with the
surrounding environment and to improve some utility function. In any system
where at least some agents are trying to innovate, the perfect strategy of
today may, with time, become a loosing one. It is the well known ``red queen
effect'': You must run as fast as you can, to stay in the same place.

It is in the economy field that innovation has been more extensively
studied. Three main types of innovations were identified:

(i) \textit{Market innovation} : the identification of new markets and
finding out how they are better served or how they may become more receptive
to the available products

(ii) \textit{Product innovation} : the identification and development of new
products

(iii) \textit{Process innovation} : the identification of better and less
expensive production ways or the improvement of internal operations

Although these classification types were developed for economics, it is an
easy exercise to find the corresponding notions in the other environments.
That also applies to the classification of the \textit{intensity} of the
innovations as \textit{radical}, \textit{incremental}, \textit{architectural}
and \textit{modular}. An important point to emphasize is that the intensity
of the innovation is an agent-dependent concept. An innovation that is
radical for one agent might just appear as incremental or of any other type
to some other agent \cite{Afuah}. Another important concept concerning a
system of innovation\cite{Padmore} is the flow of information \cite{Maslov} 
\cite{Marz} between the agents in the system and its appropriation in terms
of knowledge. However, here, systems of innovation are not explicitly taken
into account. They may only appear as emergent features.

Other important issues in the innovation field are the identification of the
basic mechanisms leading the agents to innovate \cite{Kaufmann} \cite
{Kleinknecht} \cite{Porter} \cite{Montalvo} \cite{O'Brien} \cite{Damanpour}
and its impact on social change and human evolution \cite{StPaul}.

The fact that innovation covers so many different fields and particular
settings justifies efforts to develop an abstract model that might be easily
adapted to any particular field. The dynamical structure of the model should
also be sufficiently general to provide general insights on the mechanisms
leading to emergent collective structures. In general, in an interacting
environment, the agents receive something from the environment (the other
agents) in exchange for their effort and pay the environment a certain
amount of value for the fulfilling of their needs or for the very price of
existence in that environment. We will code the two types of exchanges by
two bit strings which conventionally we denote by the \textit{products}
string and the \textit{needs} string. In an economy environment, products
and needs might be actual market products and operating or supplies needs,
but in a biology environment they might stand, for example, for hunting
success and predation by other species, in the political setting for
``slogans and promises'' and voters desires, etc.

The dynamics of the exchange is always based on the matching of the \textit{%
products} string of each agent with the \textit{needs} strings of the other
agents. Two types of models will be studied. In the first, separating the
products and needs functionalities, we study a model of producers and
consumers, the main aim being to characterize the conditions for innovation
success. In the second, each agent is equipped with two strings representing
either products and needs or, more generally, what an agent profits from the
environment and what the environment profits from him. In this model we
study how, (starting from a uniform distribution of fitness, randomly chosen
strings and equal dynamics) structure develops in the agents' society, both
with and without innovation.

\section{A model of producers and consumers}

At the start, there are $2N$ agents in the model: $N$ producers and the same
number of consumers. Each consumer has a set of \textit{needs} coded by a
string of $k$ bits and each producer has a \textit{product} coded by a
string of $k$ bits. The bit string of a consumer represents what the
consumer agent \textit{needs} to receive from the environment and the bit
string of a producer is a code for the \textit{products} that he is able to
supply. No passive actors are assumed in the environment and the environment
for each agent is just the set of all the other agents.

In addition to the two bit strings that code for needs and products, each
agent has a scalar variable $S$ or $C$, depending on the agent type
(consumer or producer, respectively). The variable $S$ represents the degree
of satisfaction of the needs and $C$ represents the amount of some commodity
(or cash) that may be exchanged for the products that are available. In the
economy this role is played by money, but in other contexts it might be
protection capacity or power or status.

The dynamics of the model is characterized by \textit{exchange}, \textit{%
evolution} and \textit{adaptation}. The basic driver of the \textit{exchange
dynamics} of the model is the matching between needs and products. At each
time step, the matching between needs and products is made and each consumer
chooses at random one among the products that better match his needs. The
producer that has this product is a potential supplier. The dynamical
evolution is

\begin{equation}
\begin{array}{l}
S_{i}(t+1)=S_{i}(t)-ac+\frac{q_{ij}^{*}}{k}\label{1.00}
\end{array}
\label{2.1}
\end{equation}

\begin{equation}
\begin{array}{l}
C_{j}(t+1)=C_{j}(t)-ap+\sum_{j(i)}\frac{q_{ij}^{*}}{k}
\end{array}
\label{2.2}
\end{equation}
The index $j(i)$ runs over all the consumers $j$ that are supplied by the
producer $i$

On receiving a product from the producer $i$ the consumer $j$ increases his
satisfaction (or \textit{energy}) $S$ by $\frac{q_{ij}^{*}}{k}-ac$. The
variable $q_{ij}^{*}$ stands for the matching of the producer $i$ that
supplies the consumer $j$. At the same time, the producer $j$ increases his
commodity (or \textit{cash}) $C$ by $\sum_{j(i)}\frac{q_{ij}^{*}}{k}-ap$,
where $ac$ and $ap$ stand for two constant \textit{costs of living} that are
subtracted at each time step from the consumers-satisfaction and the
suppliers-cash.

At each time step needs and products are compared. The producer that
supplies each consumer is chosen at random among those with the larger
matching. When $C_{i}<0$ this producer $i$ either disappears and is not
replaced (subsection 2.1) or it is replaced by a new random producer
(subsection 2.2). When $S_{j}<0$ this consumer $j$ is replaced by a new one
with random needs string and $S_{i}=S_{0}$. As such, a consumer only remains
in the field as long as its \textit{energy} $S$ is positive. If it becomes
negative, he dies and is replaced by a new random consumer. Initially all
agents and the replacement (consumer) agents are endowed with the same
initial $C_{0}$ and $S_{0}$.

The replacement mechanism of the agents means that, when applied to real
world situations, each agent in the model represents a new consumer trend
and in biology not an individual species but an ecological niche.

Once the number of surviving producers stabilizes, several possible
evolution mechanisms may be implemented in the model:

\begin{itemize}
\item  \textit{Innovation by the producers}

\begin{itemize}
\item  \textit{Market-oriented innovation}: the innovating \textbf{producers}
find the consumers that have a matching above a certain threshold and flips
the worse bit. Corresponds to adaptation of a particular product to expand
an existing market.

\item  \textit{Process innovation}: process innovation corresponding to a
decrease in production costs may be simulated in the model by adding a
certain amount (an half point for example) to the matching results of this
producer.

\item  \textit{Product} \textit{innovation: }the innovating producer finds a
set of consumers that among themselves have a matching above a certain
threshold and develops a new product string according to their need bits.
\end{itemize}

\item  \textit{Evolution} and \textit{Adaptation of the consumers}: after
the exchanges, the less satisfied \textbf{consumers} find the products that
have a matching above a certain threshold and flip the (need) bits with the
worst scores when compared with the same position bits in the products.
\end{itemize}

There are of course some important features of real world environments that
are not explicitly included in our abstract coding of the products offered
by each agent. For example, products sometimes have some core features that
are fixed and some others that are adjustable. Then the agent may supply the
same core product to different customers as different offerings. This 
\textit{market segmentation} technique is particularly important in the
services industry \cite{Gima} \cite{Johne}. The choice preference in the
model being achieved by maximization of the partial matching between
products and needs, one may take the point of view that one is dealing only
with the core features of products. An explicit coding of core versus
adjustable features might be included by keeping some product bits fixed and
fuzzying a few others. However, we believe that the qualitative dynamical
features of the model would not be very much affected by this change.

In the next subsections, the model is tested in several different scenarios,
which are characterized by different combinations of the parameter values,
namely:

\begin{itemize}
\item  The consumer cost of living parameter $ac$, providing either stable
or volatile environments

\item  The producer cost of living parameter $ap$, providing environments
with either low or high amounts of dying (or replacement) rates of producers
by time step,

\item  the innovation mechanism that is adopted: either Market-oriented
innovation (MOI) or Adaptation to available products (CAP)

\item  the quantity of agents that are allowed to perform the above
mentioned mechanisms; two possibilities have been considered: just one agent
or a randomly determined number of agents.
\end{itemize}

\subsection{Market-oriented innovation}

In the first two scenarios, the innovation mechanism is Market-oriented
innovation (MOI) by one innovating producer. Scenarios 1 and 2 differ on the 
$ac$ value, representing either stable ($ac=0.5$) or volatile ($ac=1$)
consumer environments. In both cases producers with $C_{i}<0$ are not
replaced.

In each scenario, one looks for correlations between the nature of the
market and the efficiency of the innovation process. The rate of MOI\
efficiency ($g$) of an innovating producer (IP) is defined by

\begin{equation}
\begin{array}{l}
g_{ip}=\frac{C_{ip}(t_{end})-C_{ip}(t_{innov})}{t_{end}-t_{innov}}\label%
{3.00}
\end{array}
\label{2.3}
\end{equation}
where $C_{ip}(t_{end})$ and $C_{ip}(t_{innov})$ represent, respectively, the
amount of Cash of the innovating producer at the end of the simulation and
when innovation starts.

\subsubsection{Stable and volatile environments with one innovating agent}

\begin{figure}[htb]
\begin{center}
\psfig{figure=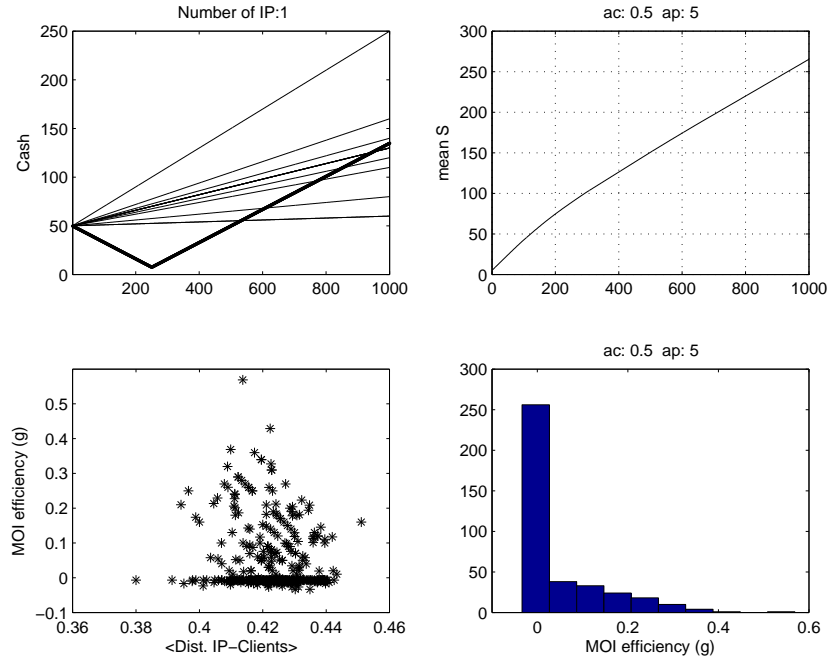,width=11truecm}
\end{center}
\caption{Stable environment and one innovating producer}
\end{figure}

\begin{figure}[htb]
\begin{center}
\psfig{figure=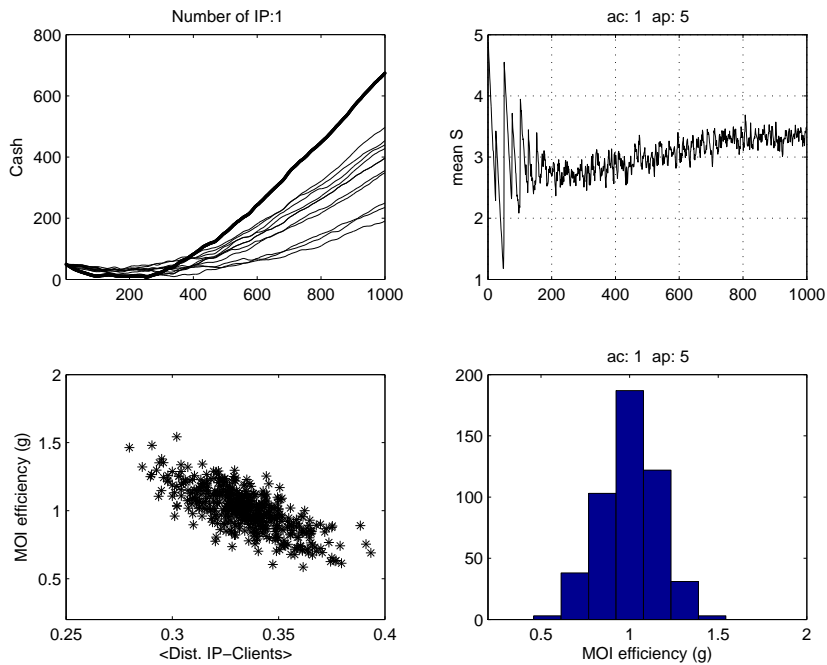,width=11truecm}
\end{center}
\caption{Volatile environment and one innovating producer}
\end{figure}

With $t_{innov}=250$ and $t_{end}=1000$, some results are shown in Figs. 1
and 2 for $ac=0.5$ and $ap=5$. At time $t_{innov}$ the surviving producer
with the lowest cash starts the market-oriented innovation process as
defined above. The two upper plots in the figures show the cash evolution
and the mean consumer satisfaction for a typical run. The bold line refers
to the innovating producer. The lower plots are obtained with a large number
of runs. The histograms in the lower right plots show that this type of
innovation is much more efficient on a highly volatile environment ($ac=1$)
than in a stable consumer environment ($ac=0.5$).

We also found inverse correlation between MOI efficiency and the distance of
the innovating producer ($IP$) to the nearest consumers (clients). But the correlation
is strong only in a volatile environment, as the lower left plot in Fig 2
shows. Simulations have also shown a negative correlation of the innovation
efficiency ($g$) with the rate of gain before innovation and with the
distance to the nearest competitor.

\subsubsection{Stable and volatile environments with many innovating agents}

Here the system was tested for different numbers of innovating producers ($%
IP $). When more than one producer is allowed to innovate, the innovation
efficiency rate ($g$) is computed as the average value obtained for the set
of innovating producers. In each simulation, the number of $IP^{\prime }$s
is determined at random, being the innovating producers chosen among the
poorest ones.

\begin{figure}[htb]
\begin{center}
\psfig{figure=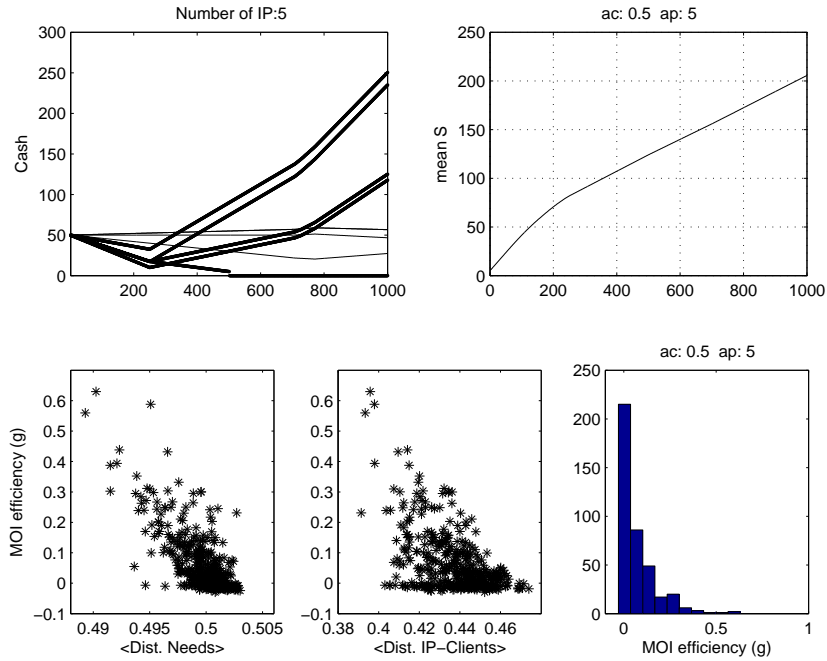,width=11truecm}
\end{center}
\caption{Stable environment and a random number of innovating producers}
\end{figure}

\begin{figure}[htb]
\begin{center}
\psfig{figure=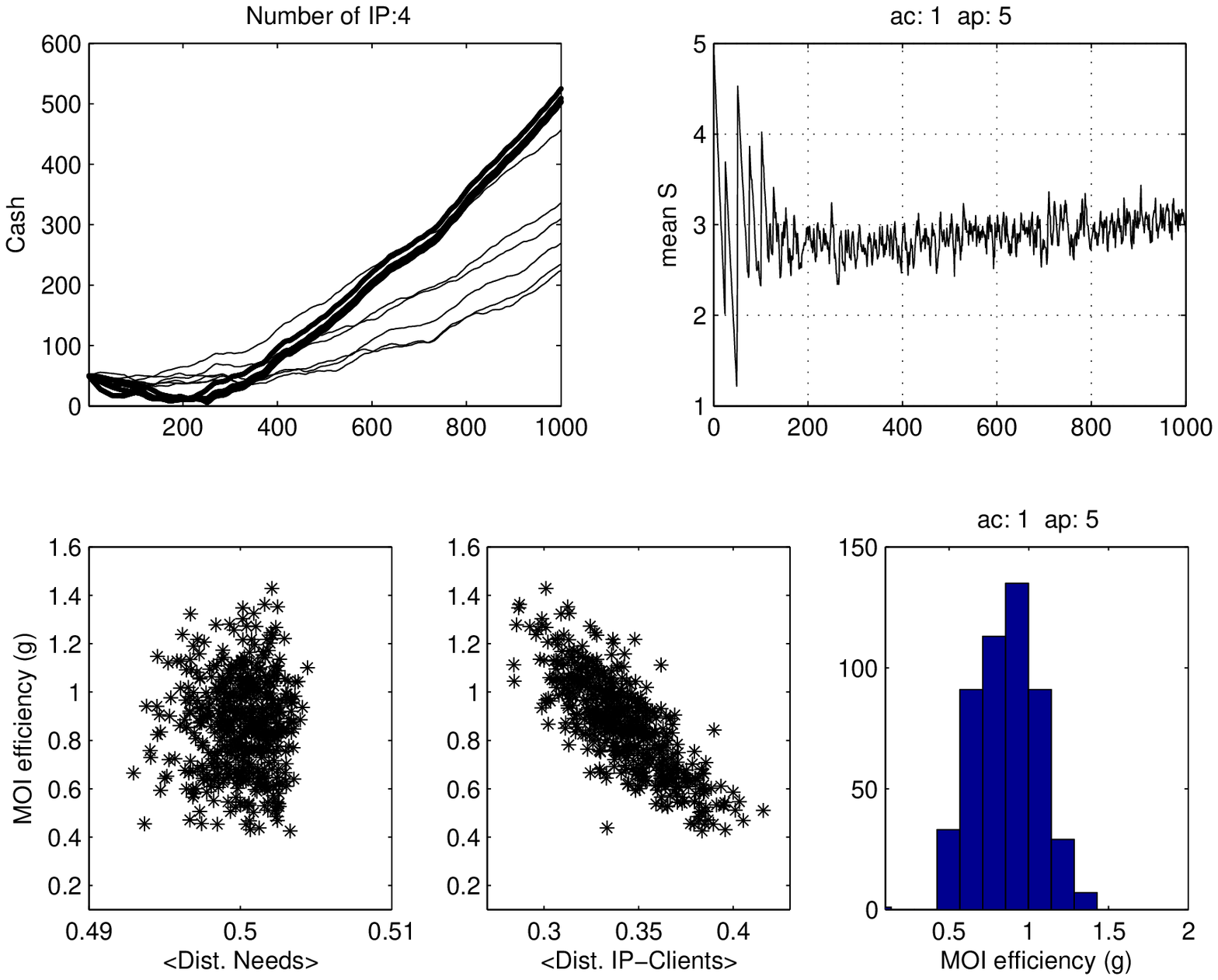,width=11truecm}
\end{center}
\caption{Volatile environment and a random number of innovating producers}
\end{figure}

The results presented in the histograms of Figs. 3 and 4 show that
Market-oriented innovation by more than one $IP$ is also much more efficient
in a volatile environment than in a stable one. On the relation between
innovation efficiency and the structure of the environment, we found inverse
correlation between MOI efficiency and the distance among consumer needs in
the stable environments. The efficiency of this type of innovation is also
inversely correlated to the distance among the innovating producers and
their nearest consumers, but the correlation is strong only in volatile
environments as the lower left plots in Fig. 4 show.

\subsection{Evolution of needs and adaptation to available products}

In this scenario, the model contains a mechanism for the \textit{evolution
of the needs}. This mechanism is one of partial adaptation or conformity
with the available products (CAP). In this scenario producers and consumers
with negative cash or satisfaction are replaced by new random ones.

\subsubsection{One adapting consumer}

When just one consumer is allowed to adapt, the model is implemented as
follows: at each time step (after the exchanges) the less satisfied consumer
finds the products that have a matching above a certain threshold and flips
the (need) bit with the worst score when compared with the same position bit
in the products.

The system is tested for different values of $ap$, allowing to simulate
environments with different rates of replacement of producers. In this
scenario, a producer only remains in the field as long as its capital ($C$)
is positive; if it becomes negative, the producer is replaced by a new
random producer.

A new efficiency rate ($s$) is then defined in order to compute the
difference between the Satisfaction level of the innovating consumer ($IC$)
after and before innovation,

\begin{equation}
\begin{array}{l}
s_{ic}=\frac{S_{ic}(t_{end})-S_{ic}(t_{innov})}{t_{end}-t_{innov}}
\end{array}
\label{2.4}
\end{equation}
where $S_{ic}(t_{end})$ and $S_{ic}(t_{innov})$ represent, respectively, the
amount of Satisfaction (or energy) of the innovating consumer at the end of
the simulation and at the start of the innovation process.

\begin{figure}[htb]
\begin{center}
\psfig{figure=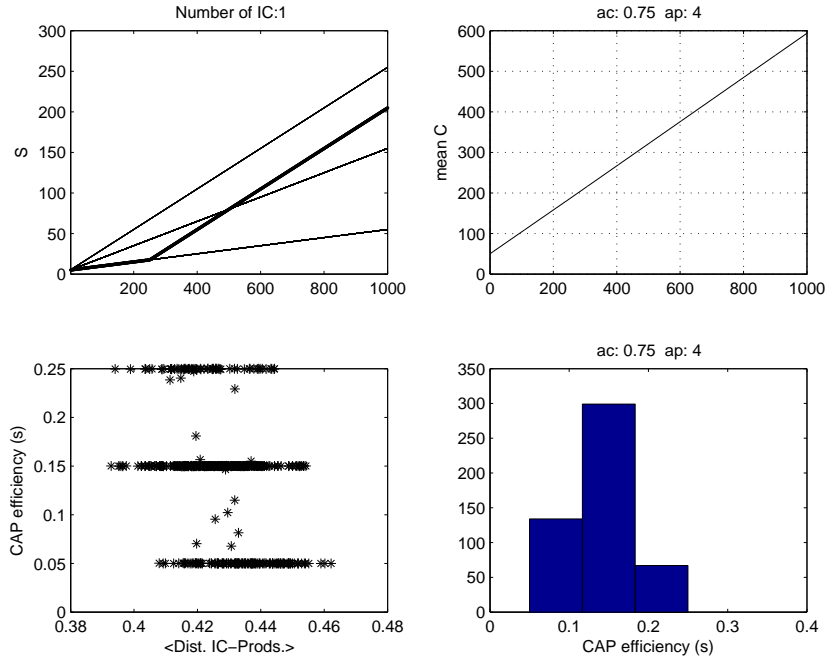,width=11truecm}
\end{center}
\caption{One innovating consumer in an environment with a low rate of
replacement of products}
\end{figure}

\begin{figure}[htb]
\begin{center}
\psfig{figure=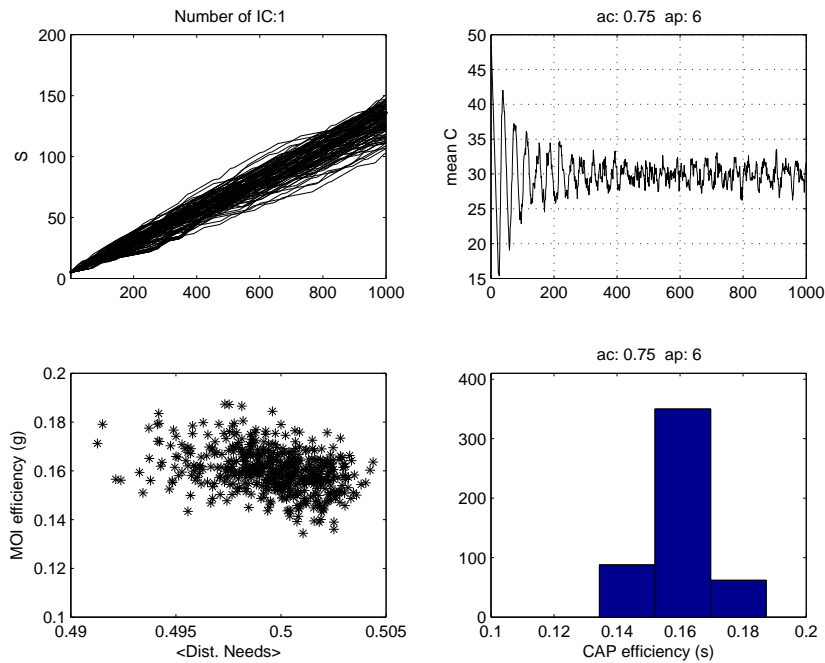,width=11truecm}
\end{center}
\caption{One innovating consumer in an environment with a high rate of
replacement of products}
\end{figure}

The results presented in the histograms of Figs.5 and 6 show that
adaptation to the available products is equally efficient either for a small
or a large rate of replacement of producers ($ap=4$ or $ap=6$, respectively).

We found inverse correlations between the satisfaction rate ($s$) and the
distance between the $IC$ and the overall set of producers in environments
with a low rate of substitution of producers, as the lower left plot in Fig.5
 shows. From the lower left plot in Fig.6 we see that when $ap$ increases,
there is a correlation between the satisfaction rate and the average
distance between needs.

\subsubsection{Many adapting consumers}

Here, the model is tested with different quantities of innovating consumers.
When more than one consumer is allowed to adapt, the satisfaction rate $s$
is computed as the average of the satisfaction rate for the set of the
innovating consumers.

\begin{figure}[htb]
\begin{center}
\psfig{figure=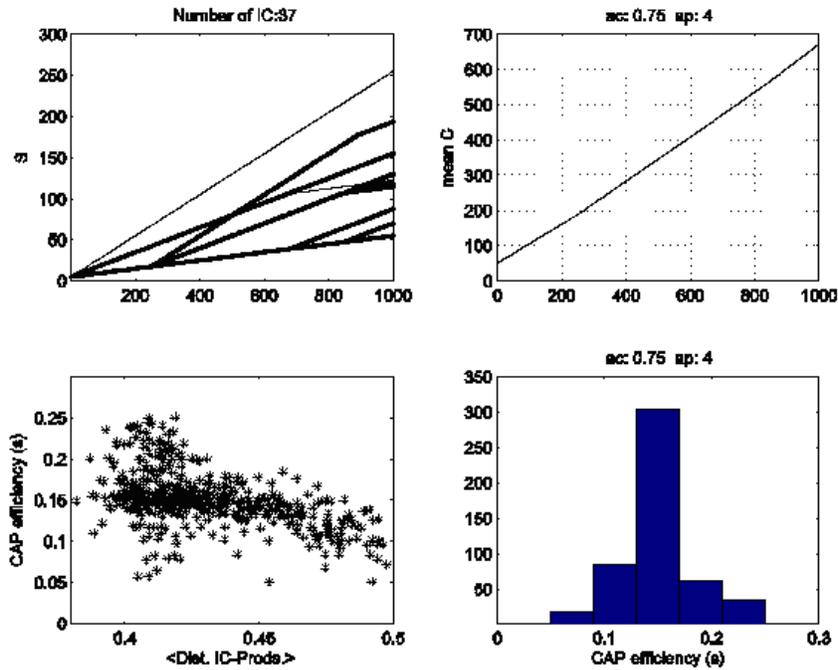,width=11truecm}
\end{center}
\caption{Several innovating consumers in an environment with low rate of
replacement of products}
\end{figure}

\begin{figure}[htb]
\begin{center}
\psfig{figure=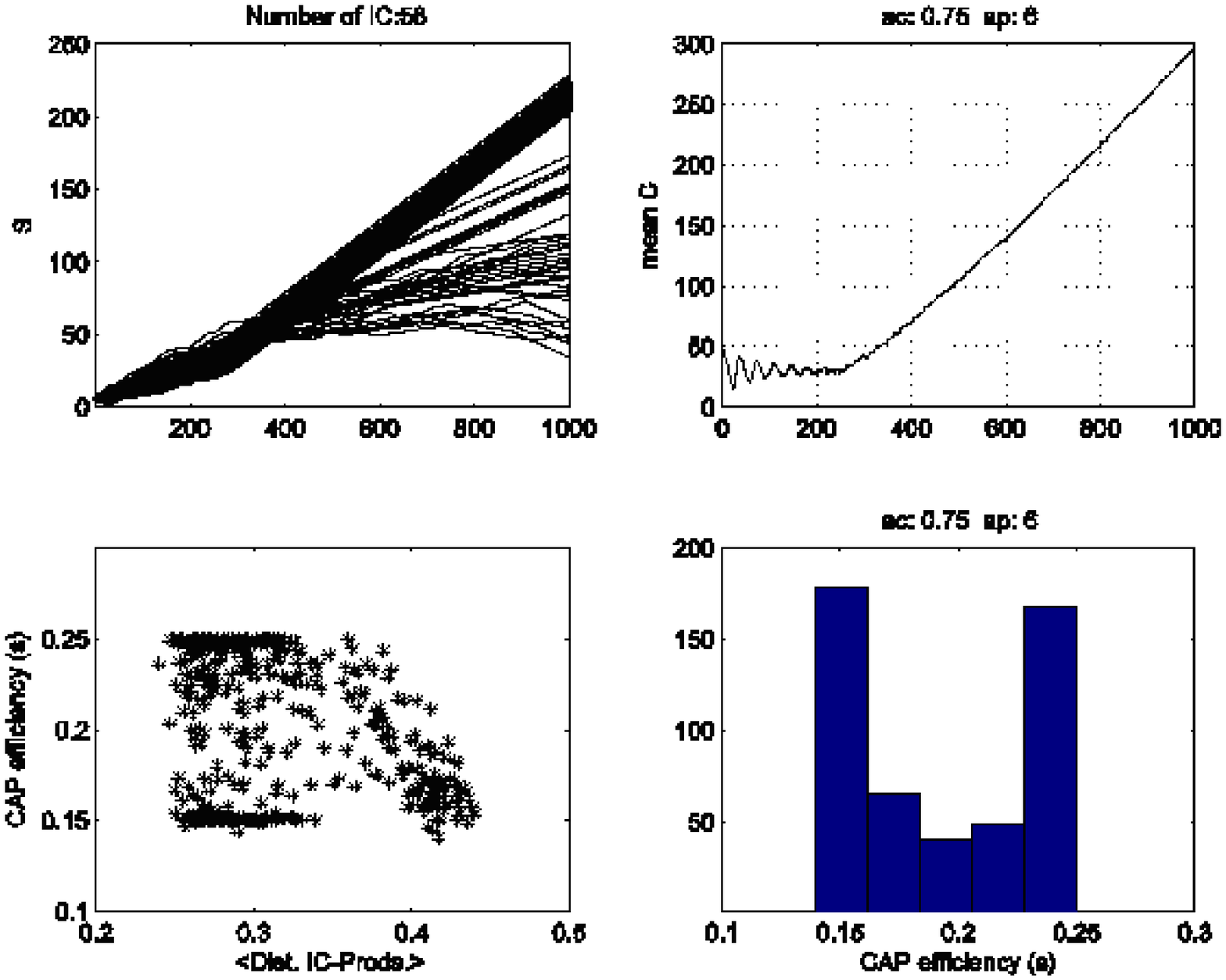,width=11truecm}
\end{center}
\caption{Several innovating consumers in an environment with high rate of
replacement of products}
\end{figure}

The results presented in the histograms of Figs. 7 and 8 show that
adaptation to the available products by many innovating consumers is
slightly more efficient in a market with a high rate of substitution of
producers $(ap=6)$ than in the case where the rate of replacement is low ($%
ap=4$).

The third plot in Fig. 7 shows a weak inverse correlation between the
satisfaction rate and the distance between the $IC^{\prime }$s and the
overall set of producers, in environments with a low rate of substitution of
producers. When the environment has a high rate of substitution of
producers, the third plot in Fig. 8 shows a weak negative correlation of the
satisfaction rate with the distance between the $IC^{\prime }$s and the
producers.

\section{A self-organizing agent model: Innovation and emergent structures}

Here we study a model where all agents have two strings, which as before we
denote as the $P$ string and the $N$ string. Here however, rather than
products and needs, as in the producers and consumers model, it is more
appropriate to interpret the $P$ string as the code for the benefits (or
energy) that the agent can extract from the environment (the other agents)
and the $N$ string as a code for what the other agents may extract from him.

As before, the dynamical evolution is based on the matching of the $P$ and $%
N $ strings. Each agent has a fitness function $F$ which evolves as follows 
\begin{equation}
F_{i}\left( t+1\right) =F_{i}\left( t\right) +\sum_{j(i)}\frac{q_{ij}^{*}}{k}%
-\frac{q_{l\left( i\right) i}^{*}}{k}  \label{3.1}
\end{equation}
$\sum_{j(i)}$ denotes a sum over all the agents $j$ for which the $P$ string
of $i$ has maximal matching $q_{ij}^{*}$. As before only one agent is chosen
at random among those with maximal matching. $q_{l\left( i\right) i}^{*}$
denotes the maximal matching of the $N$ string of agent $i$ with the other
agents. At time zero the $P$ and $N$ strings of all agents are chosen at
random and the fitness is initialized to some fixed value $F\left( 0\right) $%
. Whenever, during the time evolution, the fitness of one agent becomes
negative, this agent is replaced by another random agent with the initial
fitness.

One of the purposes of the study of this model is to show how, starting from
a set of agents in identical conditions, the time evolution spontaneously
creates fitness inequalities among them. How the structures are affected by
innovation will also be studied. In the question of creation of structure in
agent societies, an important issue is also how the evolution affects
diversity.

Innovation in the model of this section is also an adaptation of the $P$
string to the $N$ strings. Two kinds of innovation are considered. In first
(called $P-$innovation) each agent compares his $P$ string to the $N$
strings of the other agents having matchings above a threshold ($thrs$) and
flips his worst scoring bit. Therefore $P-$innovation means that the agent
tries to maximize what he receives from the other agents. In the second kind
of innovation (called $N-$innovation) each agent tries to minimize the
matching of his $N$ string with the $P$ strings of the other agents. At each
time step, this is also done by flipping a bit, this time the bit that has
the better matching. The meaning of $N-$innovation is that the agent tries
to give the other agents as little as possible or, in a sense, that is
trying to protect itself from the wearing out effects of the environment. In
this model, whenever innovation is implemented, all agents are allowed to
innovate, in line with the equal opportunity point of view of the model.

In the Figs. 9 to 12 we compare the situations in different scenarios. The
two upper plots compare the histograms of the initial fitness and the
fitness after $T=5000$ time steps. The middle histograms compare the
diversity of the $P$ strings at the initial time and at $T=5000$. Diversity
of the strings is characterized by the histogram of their Hamming distances.
The lower histograms contain a similar comparison for the $N$ strings. In
all cases where innovation is implemented, $thrs=1$.

\begin{figure}[htb]
\begin{center}
\psfig{figure=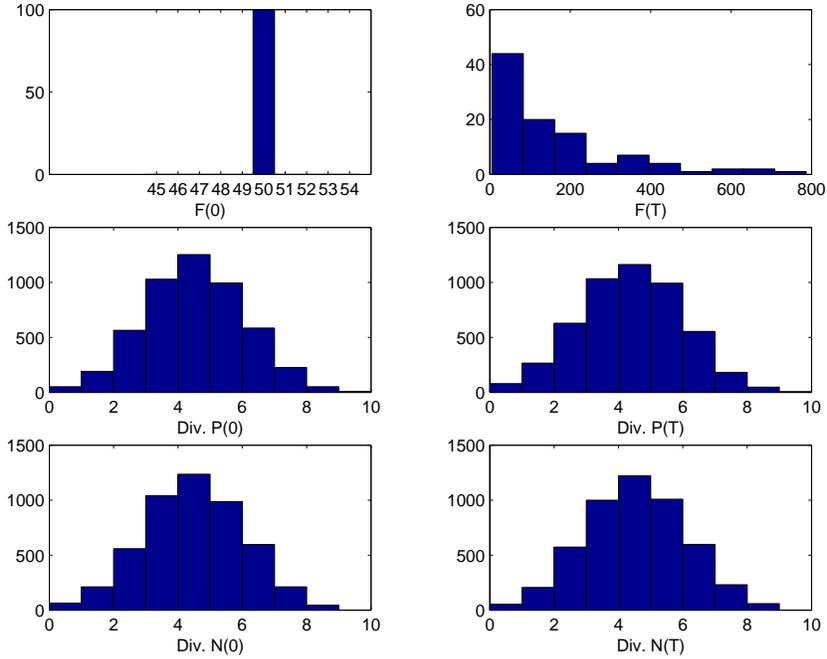,width=11truecm}
\end{center}
\caption{Fitnesses and diversity of strings without innovation}
\end{figure}

Fig. 9 is the situation without innovation. Although all agents start with
similar conditions a large stratification of fitnesses emerges as a result
of the time evolution. The model shows how a well defined structure emerges
from the dynamical evolution. Dynamics and random events generate inequality.

\begin{figure}[htb]
\begin{center}
\psfig{figure=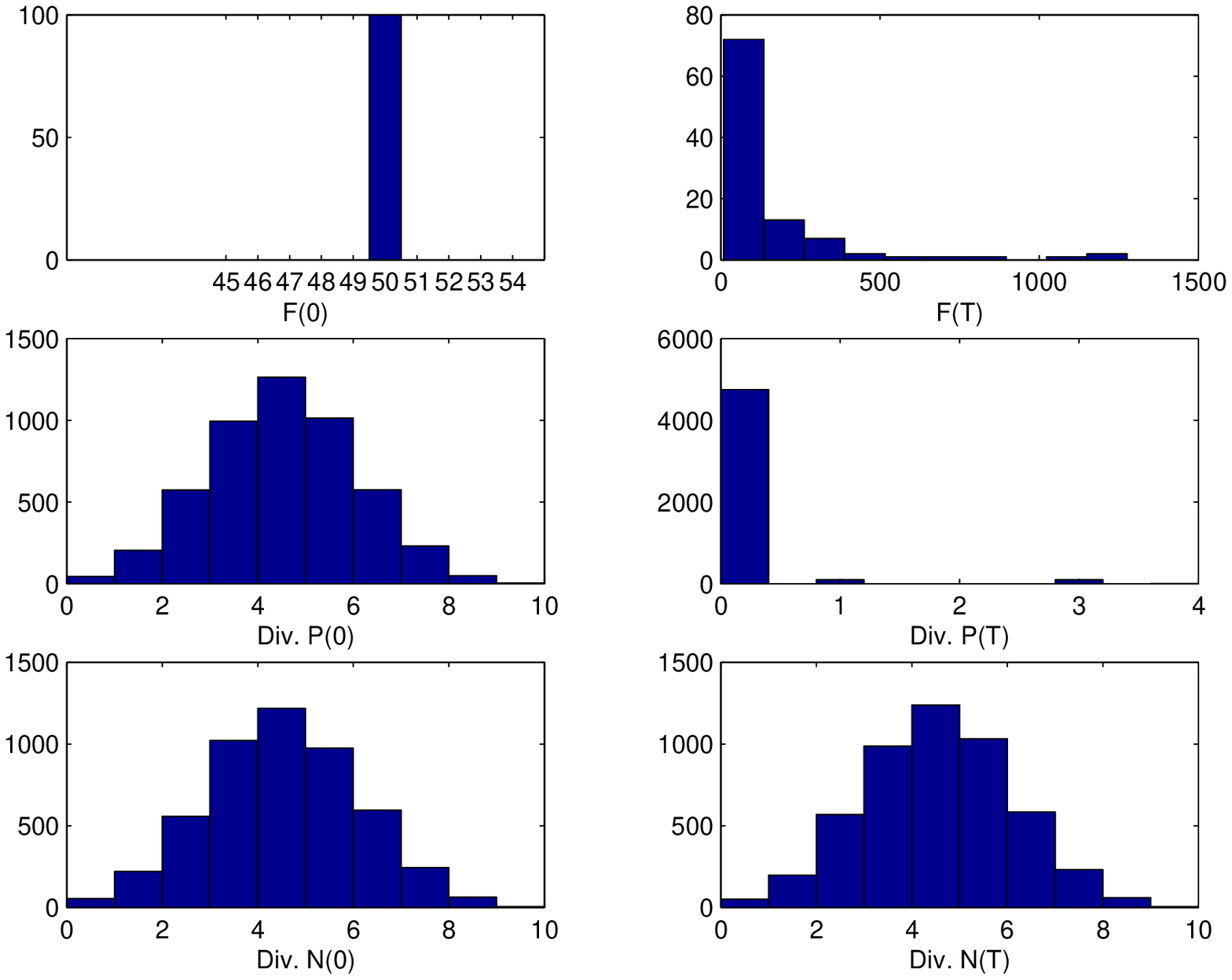,width=11truecm}
\end{center}
\caption{Fitnesses and diversity of strings with $P-$innovation}
\end{figure}

In Fig.10, $P-$ innovation for all agents is implemented. One sees that the
stratification effect is even stronger when this type of innovation is
turned on. For the dynamics without innovation the string diversities at the
initial time and at $T=5000$ are similar. However, for the $P-$innovation
case one sees a concentration of the $P$ strings around a dominant type.
Inequality stratification is enhanced and diversity decreases.

\begin{figure}[htb]
\begin{center}
\psfig{figure=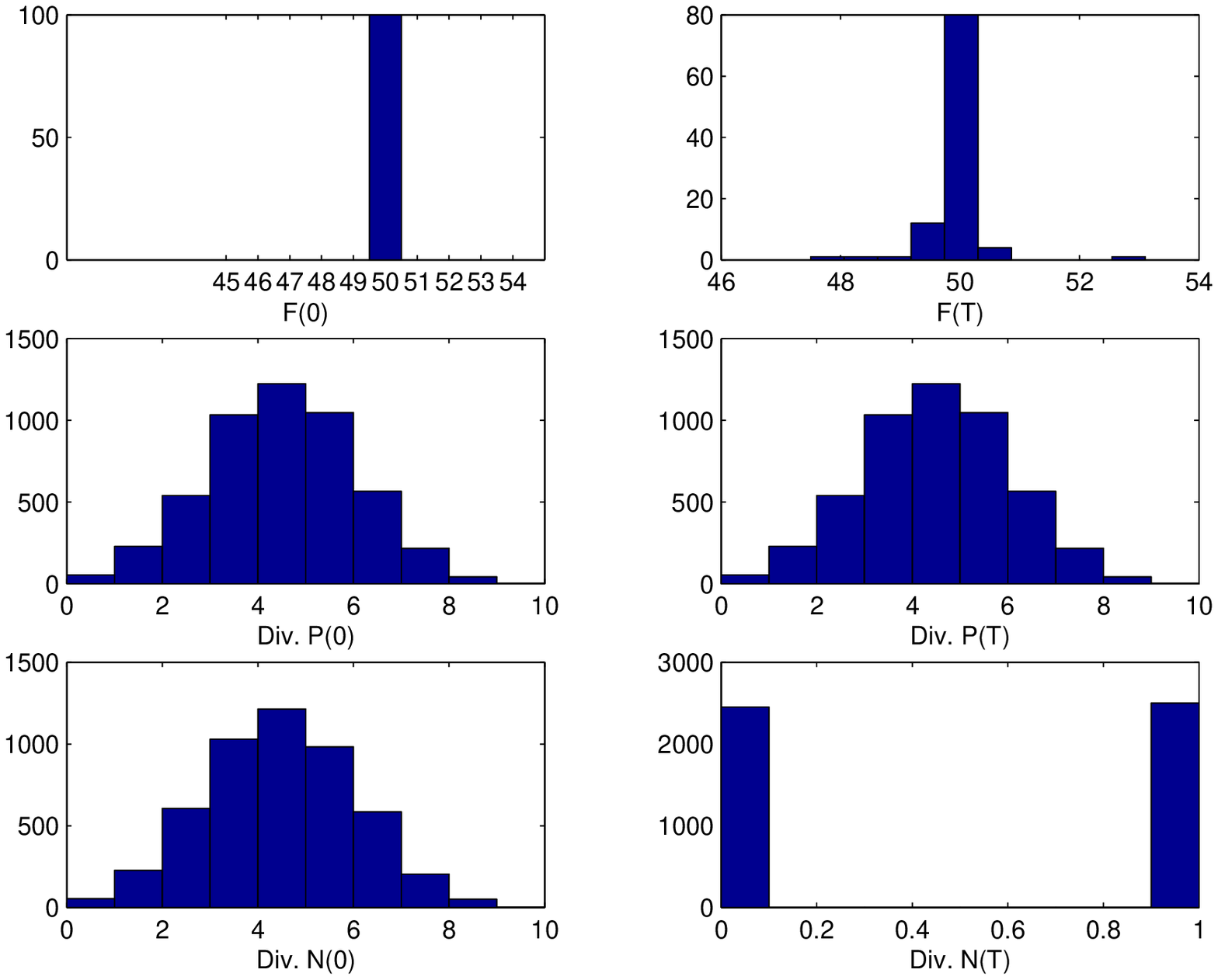,width=11truecm}
\end{center}
\caption{Fitnesses and diversity of strings with $N-$innovation}
\end{figure}

In Fig.11, all agents perform $N-$innovation. Here one sees that the final
fitnesses are not very different from the initial ones. The exchanges are
minimized, and the most relevant structure that develops is a drastic
reduction of diversity in the $N$ strings.

\begin{figure}[htb]
\begin{center}
\psfig{figure=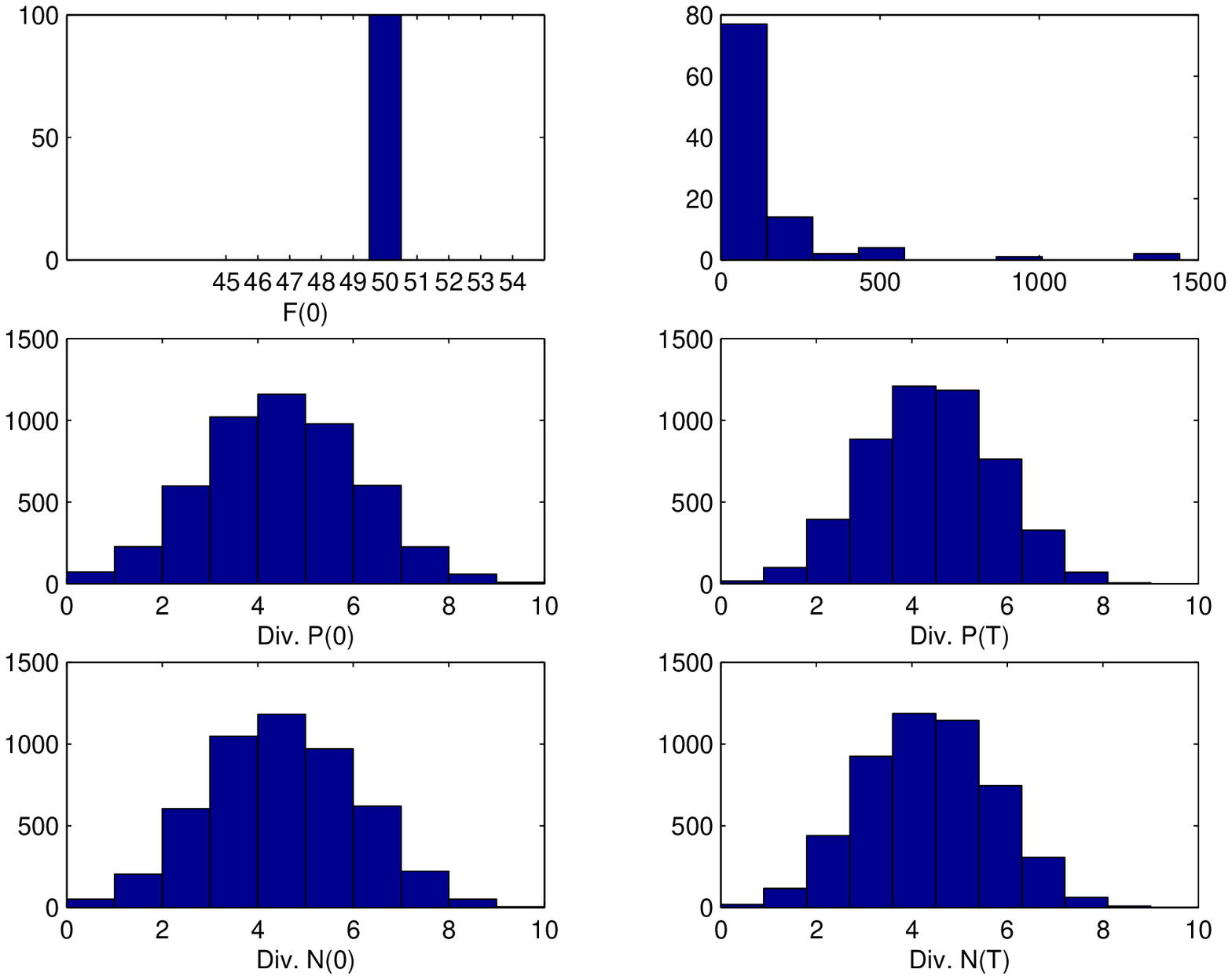,width=11truecm}
\end{center}
\caption{Fitnesses and diversity of strings with both $P-$ and $N-$%
innovation }
\end{figure}

When both $P$ and $N-$innovation are implemented (Fig.12), both the
diversity of the strings and the stratification of the fitnesses are
restored. It is interesting to notice that in terms of the global parameters
of agent's society, the two types of innovation seem to cancel out and the
results are similar to the situation without innovations.

\section{Conclusions}

A very general feature of any real world complex system is the fact that
each agent can extract something from the environment (the other passive or
active agents) and the other agents may extract something from him. This is
the basic fact behind our $P$ and $N$ coding strings and their matching.
This abstract coding allows to study general effects, independently of the
particularities of each actual complex adaptive system. In addition to the
dynamics of interaction, ruled by the matching of the strings, the actions
of the agents in their adaptation to the environment is conveniently coded
by the evolution of the bit strings.

In the first (consumers and producers) model, by separating the
functionalities associated to the $P$ and $N$ strings, we were able to
obtain very general conclusions about the effectiveness of the innovation
mechanisms and how this effectiveness relates to the overall structure of
the agents' environment and their relation to it.

In the second model, the agents being equipped with both types of
interactions with the environment, we obtained a clear manifestation of the
fact that a simple dynamics of interaction creates strong structures in
agent societies. On the other hand, active actions by the agents to improve
their fitness create further structure and, in particular, have a strong
effect on diversity. Therefore, the onset of these (innovation) actions at a
particular time may be the driving mechanism for structural changes in
actual real world situations.

\end{document}